\documentclass[aps,pra,showpacs,twocolumn,superscriptaddress]{revtex4}

\usepackage{graphicx}
\usepackage{dcolumn}
\usepackage{bm}
\usepackage{amsfonts}
\usepackage{amssymb}
\usepackage{amsmath}
\usepackage{bbm}
\usepackage{color}

\newcommand{\ket}[1]{|#1\rangle}
\newcommand{\bra}[1]{\langle #1|}
\newcommand{\braket}[2]{\langle #1|#2 \rangle}
\newcommand{\beq}{\begin{equation}}
\newcommand{\eeq}{\end{equation}}
\newcommand{\bea}[1]{\begin{equation}\begin{array}{#1}}
\newcommand{\eea}{\end{array}\end{equation}}
\newcommand{\beqn}{\begin{eqnarray}}
\newcommand{\eeqn}{\end{eqnarray}}

\providecommand{\openone}{\mathbbm{1}}
\usepackage{color}

\date{\today}


\begin{document}

\title{Detection and typicality of bound entangled states}

\author{Joonwoo Bae}
\affiliation{School of Computational Sciences, Korea Institute for Advanced Study, Seoul 130-012, Korea}

\author{Markus Tiersch}
\affiliation{Physikalisches Institut, Albert-Ludwigs-Universit\"at, Hermann-Herder-Str.~3, D-79104 Freiburg, Germany}

\author{Simeon Sauer}
\affiliation{Physikalisch-Astronomische Fakult\"at, Friedrich-Schiller-Universit\"at Jena, Max-Wien-Platz 1, D-07743 Jena, Germany}

\author{Fernando de Melo}
\email{fernando.demelo@physik.uni-freiburg.de}
\affiliation{Physikalisches Institut, Albert-Ludwigs-Universit\"at, Hermann-Herder-Str.~3, D-79104 Freiburg, Germany}

\author{Florian Mintert}
\affiliation{Physikalisches Institut, Albert-Ludwigs-Universit\"at, Hermann-Herder-Str.~3, D-79104 Freiburg, Germany}

\author{Beatrix Hiesmayr}
\affiliation{Faculty of Physics, University of Vienna, Boltzmanngasse 5, A-1090 Vienna, Austria}

\author{Andreas Buchleitner}
\affiliation{Physikalisches Institut, Albert-Ludwigs-Universit\"at, Hermann-Herder-Str.~3, D-79104 Freiburg, Germany}

%

\begin{abstract}
We derive an explicit analytic estimate for the entanglement of a large class of bipartite quantum states which extends into bound entanglement regions. This is done by using an efficiently computable concurrence lower bound, which is further employed to numerically construct a volume of  $3 \times 3$ bound entangled states.

\end{abstract}

\maketitle

%

It is one of the most challenging and
fundamental issues in quantum information 
science to decide whether a given quantum state 
can exhibit quantum correlations, i.e., whether it is {\em entangled}. 
This question is fundamental inasmuch as it rephrases the quest for 
the quantum-classical demarcation line, and it is also of potentially enormous practical 
relevance -- in view of the many applications of quantum theory in modern 
information technology. Since entanglement is fragile, hard to screen against 
the detrimental influence of decoherence and rapidly reduced to 
a residual level under environment coupling, it is important to realize that 
even quantum states which are ``close" to separable 
states and, in this sense, carry only residual amounts of entanglement, still might be used to accomplish typical tasks 
of quantum information processing, after ``distillation"~\cite{bennett:722}: Many weakly entangled states can be processed to condense their 
collective entanglement content in one strongly 
entangled state, which then can be used to solve the predefined task.

However, there are entangled states from which no entanglement can be distilled, accordingly called {\em bound entangled states}~\cite{boundent}.  
If, e.g., two parties were to set up a quantum channel by sharing an entangled state, environmental noise can escort that state to a bound entangled state, thus
preventing any subsequent distillation -- the quantum communication channel will be ill-fated.  An entangled state $\rho$ that is positive under partial transpose (PPT) -- i.e. $(\openone\otimes T)(\rho_{ij,kl}) =\rho_{il,kj}\ge0$ --  was shown to be bound entangled~\cite{ppt,horodecki:333,nota01}.  In other words, the only easily computable entanglement measure to date~\cite{vidalNeg} fails to detect exactly those states that represent a severe problem for quantum communication protocols. It is therefore mandatory to develop tools to efficiently map out the volume of bound entangled states, which is hitherto
barely characterized: So far, only continuous families of optimal entanglement witnesses could be used to delimit bound entangled states~\cite{lewenstein:052310,bertlmann:052331,baumgartner:032327,bertlmann:024303,bertlmann:014303}, by 
intersecting the volume of entangled states detected by the witnesses with the volume of quantum states with positive partial transpose. All quantum states within this intersection are bound entangled. The crux of this method lies in the difficulty of constructing optimal witnesses, which is known to be a computationally hard task, in general. Here we show that an  algebraic lower bound of entanglement when quantified by concurrence -- which can be evaluated analytically or by numerical diagonalization -- can be employed for efficient detection of an important class of bound entangled states of finite dimensional bipartite quantum systems.   

%

Let us start by a short recollection of the basic definitions of concurrence and its lower bound as
employed hereafter. As 
shown elsewhere~\cite{mintert:260502,mintert:207}, Wootter's original concurrence definition for pure states~\cite{hills97} can be 
re-expressed in terms of the expectation value of a projector-valued operator  ${\cal A}$,
\beq {\mathcal{C}}\left(\Psi \right) = \langle \Psi | \otimes
\langle \Psi| {\cal A} | \Psi\rangle\otimes | \Psi\rangle ^{1/2}\, ,
\label{cPure} 
\eeq
where ${\cal A}$ acts simultaneously onto two versions of the state:
\begin{equation}
 \mathcal{A}=4 \sum_{i<j,k<l} \big(
 \ket{ikjl}-\ket{jkil}-\ket{iljk}+\ket{jlik}
 \big) \cdot (h.c.)
\end{equation}
Here, $i$ and $j$ enumerate the basis vectors of the first partition, and $k$ and $l$ the second partition's.
This definition of concurrence can be generalized for mixed states $\rho$, as an infimum over all possible pure state decompositions defined in terms of probabilities $p_i$ and pure states $| \Psi_{i}\rangle$:
\beqn
{\mathcal C}\left(\rho\right) & = & \inf_{\left\{
p_{i},|\Psi_{i}\rangle\right\} }\sum_i p_{i} \langle \Psi_{i} |  \otimes
\langle \Psi_{i}| {\cal A} | \Psi_{i}\rangle \otimes | \Psi_{i}\rangle
^{1/2}\, . 
\label{cr} 
\eeqn
This latter optimization problem has an explicit algebraic solution for pairs of
qubits~\cite{wootters:2245}, but admits only numerical solutions or algebraic estimates if the system size is increased 
-- either by an increase of the constituents' sub-dimension, or of their number. 

In the following,
we will use a specific, algebraic estimate, the {\em quasi pure} lower bound, which is 
easily evaluated (by 
diagonalization of a matrix of the same dimension as $\rho$) and known 
to yield good estimates for weakly mixed states~\cite{mintert:012336}. In short, it is obtained from the singular values 
$\mathcal{S}_{i}$ of a matrix 
with elements
\begin{equation}
\mathcal{T}_{ij}=\sqrt{\mu_i\mu_j}\bra{\Psi_i}\otimes\braket{\Psi_j}{\chi}\, ,
\label{eq:T_alpha}
\end{equation}
that can easily be constructed with the spectral decomposition
$\rho=\sum_i\mu_i\ket{\Psi_i}\bra{\Psi_i}$, and choosing 
$\ket{\chi}\propto {\cal A} \ket{
\Psi_0} \otimes \ket{\Psi_0}$, with $\ket{\Psi_0}$ the
dominant eigenvector of $\rho$
(associated with the density matrix' largest eigenvalue).
The concurrence can then be bounded from below by 
\begin{equation}
\mathcal{C}\left(\rho\right)\ge
\mathcal{C}_{\rm qp}\left(\rho\right)=\max\left(0\text{,
}\mathcal{S}_{0}-\sum_{i>0}\mathcal{S}_{i}\right)\, ,
\label{eq:lower_bound}
\end{equation}
for arbitrary states $\rho$ (in contrast to witnesses, which need to be tailored for the detection
of specific states). Therein, $\mathcal{S}_{0}$ denotes the largest singular value of matrix $\mathcal{T}$. We will employ this quasi pure lower bound $\mathcal{C}_{\rm qp}$ throughout the sequel
of this paper.

%

We now set out for identifying a volume of bound entangled states within the set 
of $d$ dimensional {\em Bell-diagonal states}, a class of states of special importance for quantum key distribution protocols \cite{cerf:127902}, and whose structure is not completely  known~\cite{vollbrecht}.
These are given as convex sums of maximally entangled Bell-like states, 
\beq \label{exp}
\rho = \sum_{k,l=0}^{d-1}
\lambda_{kl} P_{kl}
\,,
\eeq 
with probabilities $\lambda_{kl}\ge 0$,
$\sum_{k,l}\lambda_{kl}=1$, and $P_{kl} =
\ket{\Omega_{kl}}\bra{\Omega_{kl}}$ the projectors onto the Bell
states
\beq
\ket{\Omega_{kl}} =
\frac{1}{\sqrt{d}} \sum_{s=0}^{d-1} e^{\frac{2\pi i }{d} sk}
\ket{s}\ket{s+l}
\,.
\eeq
The latter are transformed into each other by local unitary operations, e. g. by the
Weyl operators $W_{kl} = \sum_{s} e^{2\pi i sk /d }
|s\rangle\langle s+l|$, such that $\ket{\Omega_{kl}} =
(W_{kl}\otimes \openone)\ket{\Omega_{00}}$. 

From several copies of the Bell state
$\ket{\Omega_{00}}$, a Bell-diagonal state is generated by
introducing simple errors such as phase-shifts and
bit-translations, and error correction in general allows to reverse this process,
and thus to distill a maximally entangled Bell state from a ``reservoir'' of Bell diagonal 
states. Nonetheless, bound entangled Bell-diagonal states which 
{\em do not} admit entanglement distillation do exist~\cite{stormer}, and, as we will show in the following, can be effortlessly detected.  

For this purpose, we first derive  $\mathcal{C}_{\rm qp}$ for arbitrary Bell diagonal states as defined in (\ref{exp}). 
To do so, suppose that $P_{n m}$ has the largest weight in (\ref{exp}), i.e., that $\rho$ be quasi pure  
with respect to $P_{n m}$. We then can construct the matrix $\mathcal T$ in (\ref{eq:T_alpha}) with the
choice $\ket{\chi} \propto {\cal A} \ket{\Omega_{n m}}\otimes\ket{\Omega_{n m}}$, and the singular values 
of $\mathcal T$ are given by the square roots of the eigenvalues of ${\cal T}{\cal T}^{\dagger}$ -- which itself can be shown to be a Bell diagonal matrix.
Consequently, the singular values can be readily read off from ${\mathcal S}_{kl}^{n m} = \bra{\Omega_{kl}}{\cal T}_{n m}{\cal
T}_{n m}^{\dagger}\ket{\Omega_{kl}}^{1/2}$, with the explicit expression
\begin{widetext} 
\beq
\mathcal{S}_{kl}^{n m} = \left\{\frac{d}{2(d-1)} \lambda_{kl}\left[ \left(
1- \frac{2}{d}\right )\lambda_{n m}\delta_{k,n}\delta_{l,m} +
\frac{1}{d^2} \lambda_{(2n-k)(\text{mod} ~d),(2m-l)(\text{mod}~d)} \right]\right\}^{1/2}\, ,
\label{log}
\eeq
\end{widetext}
which can be plugged into (\ref{eq:lower_bound}) to obtain the desired result (note that the singular values 
now carry four indices: the two upper-indices refer to the Bell state with largest eigenvalue, and the two lower indices are the labels of the Bell basis). This represents the first analytical estimation of concurrence for a family of states that encompasses bound entangled states.

%

\begin{figure*}
\includegraphics[width=0.4\textwidth]{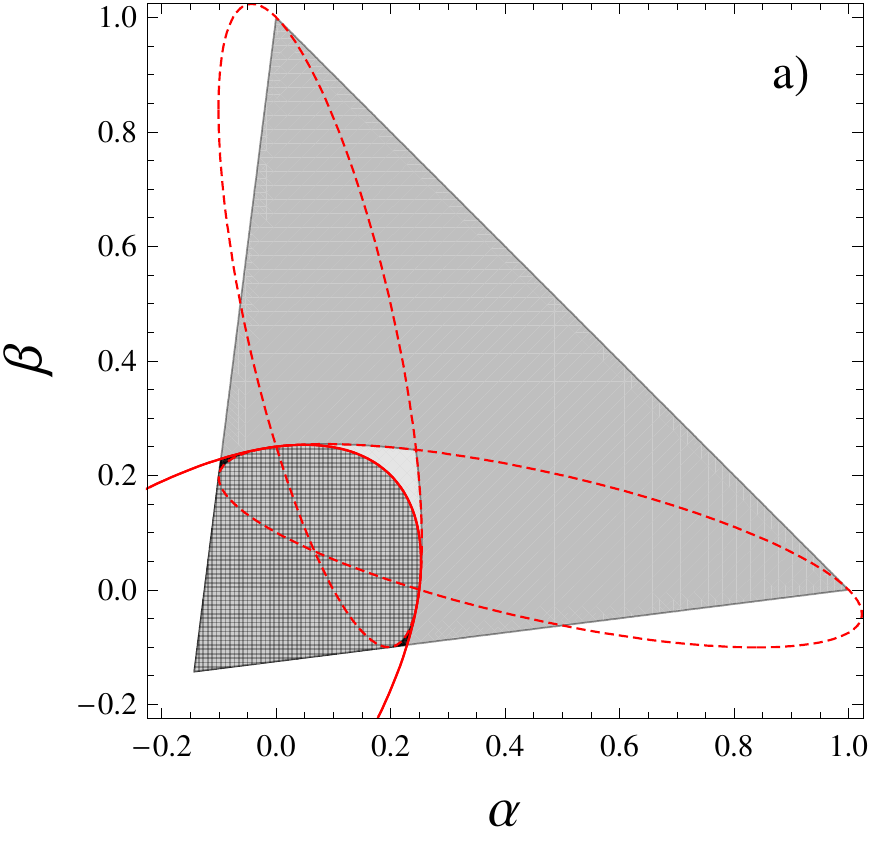}
\includegraphics[width=0.41\textwidth]{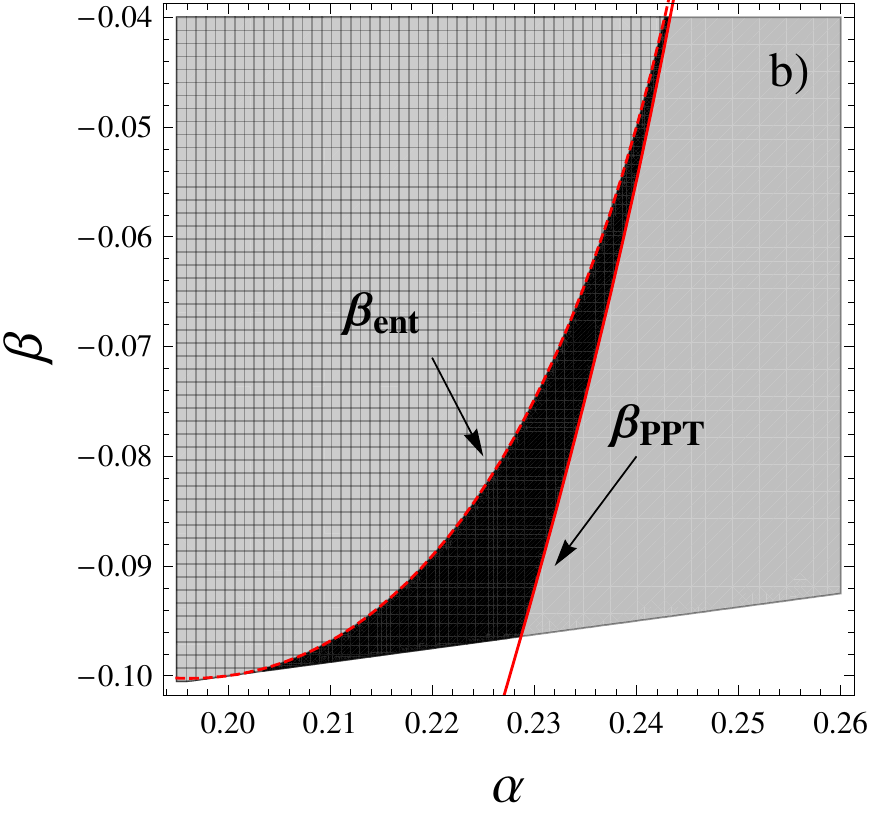}
\caption{(Color online) (a) Parameter space of the ``line'' states, as defined in (\ref{online}), for $\gamma =0$: The crosshatched dark gray area depicts states with positive partial transpose, while the quasi pure lower bound $\mathcal{C}_{\rm qp}$, Eq.~(\ref{eq:lower_bound}), vanishes in the light gray area. The 
black area corresponds to states that have positive partial transpose and $\mathcal{C}_{\rm qp}$ exhibit positive values, and thus is associated with  {\em bound entangled states}.  The solid gray triangle encompasses the positivity region ($\rho > 0$). The curves refer to the borderlines in Eq.~(\ref{borderlines}): The full line delimits the positive partial transposition region, and the dashed ellipses the bound entangled region, when over the PPT region.  Note that some entangled states, with negative partial transposition, have $\mathcal{C}_{\rm qp} = 0$, but this was not observed in areas of bound entanglement.    (b) Zoom into the bound entangled area  shown in (a). 
}
\label{fig1}
\end{figure*}

\begin{figure*}
\begin{center}
\includegraphics[width=1.\textwidth]{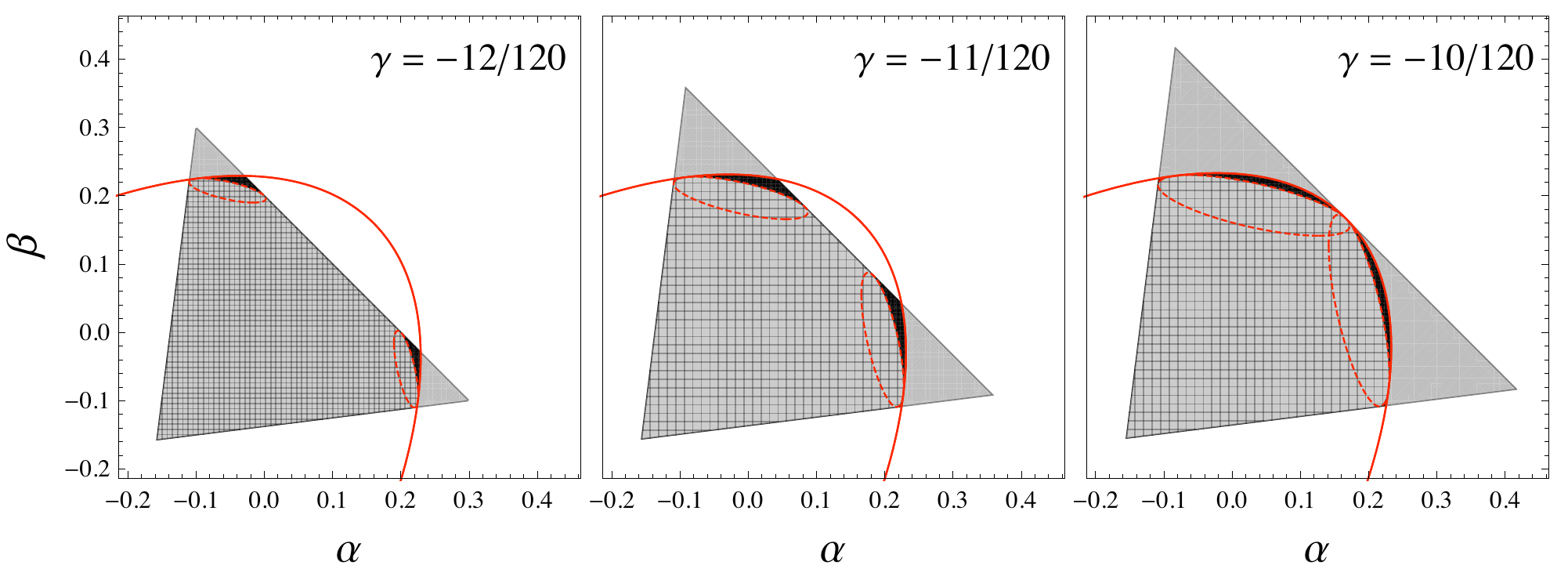}
\caption{(Color online)  Same as Fig.~\ref{fig1}, for nonvanishing values of $\gamma$ in Eq.~(\ref{online}). Also in these cases a perfect agreement is observed between the areas detected by the lower bound and the exact areas  defined by Eqs.~(\ref{borderlines}).}
\label{fig2}
\end{center}
\end{figure*}

We now apply this result to delineate the area of bound entangled states within 
the class of $3\times 3$ Bell diagonal ``line states'' defined as (see also~\cite{baumgartner:032327,beatrixPLA})
\begin{eqnarray}
\rho=\frac{1-\alpha-\beta-\gamma}{9}\mathbbm{1}+\gamma\;
P_{00}+\alpha \;P_{10}+\beta\; P_{20}\, .\label{online}
\end{eqnarray} 
(The same Weyl operator generates $\ket{\Omega_{10}}$ from $\ket{\Omega_{00}}$, as 
$\ket{\Omega_{20}}$ from $\ket{\Omega_{10}}$ -- therefore these states extend along a ``line''.)
For these states, the existence of a non-vanishing area of parameter space giving rise to bound
entangled states had already been demonstrated through the optimization of witness 
operators~\cite{baumgartner:032327,bertlmann:024303,bertlmann:014303}. With the present approach, we can effortless 
scan the entire  $\alpha$-$\beta$ plane for fixed $\gamma$. The results, for $\gamma =0$ in Fig.~\ref{fig1},
show that the intersection of the area of positive $\mathcal{C}_{\rm qp}$ with the area of positive partial transpose 
perfectly reproduces the area identified by the witness approach. 
Also note that the quasi pure bound here provides fully reliable information, despite the fact  
that the  identified bound entangled states are rather mixed (with purities somewhere around $0.17$, and a minimum
value $1/9\simeq 0.111$). Fig.~\ref{fig2} shows analogous results for $\gamma < 0$. 

As a byproduct of these results, we could exactly parametrize the borderline
of bound entangled states and of those with positive partial transpose via the witness approach~\cite{baumgartner:032327}.
For positive $\alpha$ (which suffices due to the apparent 
symmetry of parameter space as spelled out by the figures), the corresponding expressions read
\begin{eqnarray}
\beta_{\rm ent}&=&\frac{1}{8} \Big( 5 - 17\,\alpha + 19\,\gamma\nonumber\\&&
\pm 3\,{\sqrt{1 + 6\,\alpha - 39\,{\alpha}^2 + 30\,\gamma -
102\,\alpha\,\gamma +
         33\,{\gamma}^2}} \Big) \, ,\nonumber\\
\beta_{\rm PPT}&=&\frac{1}{16} \Big( -2 + 11\,\alpha +
11\,\gamma\nonumber\\&& \pm 3\,{\sqrt{4 - 12\,\alpha -
15\,{\alpha}^2 - 12\,\gamma + 66\,\alpha\,\gamma -
         15\,{\gamma}^2}} \Big) \; ,\nonumber\\
\label{borderlines}
\end{eqnarray}
and are indicated respectively by the dashed ellipses and full line in Figs.~\ref{fig1} and~\ref{fig2}: the gap region between the ellipses and the full line, inside of the PPT region, defines the bound entangled area.

\begin{figure}[t!f]
\includegraphics[width=\columnwidth]{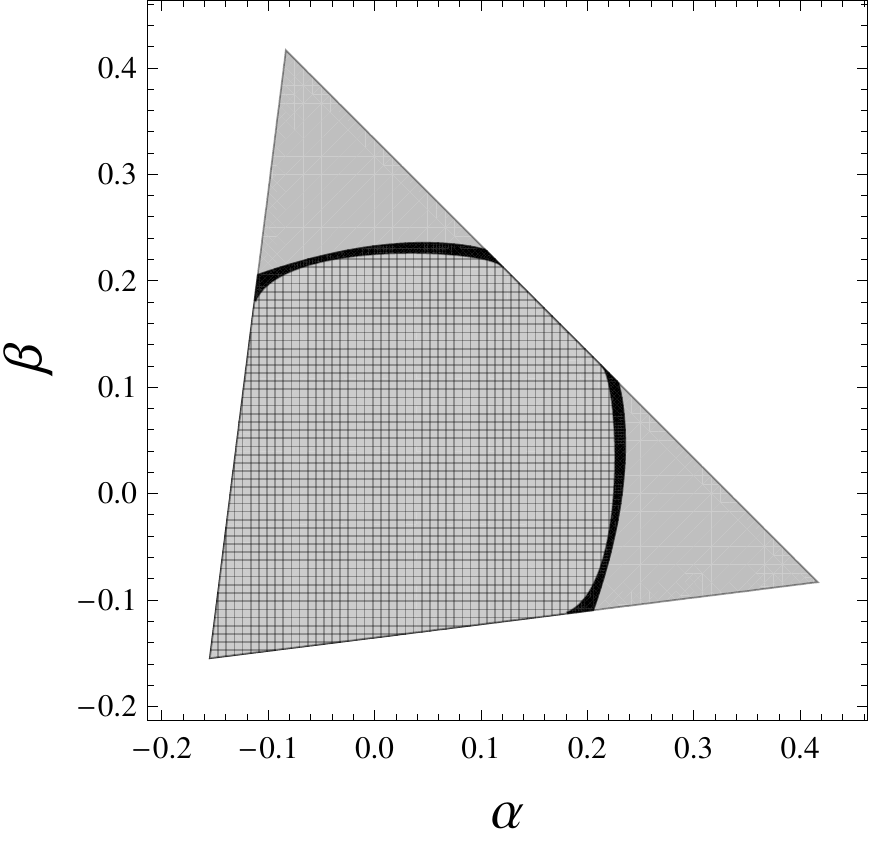}
\caption{ Parameter space of the ``beyond line" states as defined in Eq.~(\ref{beyondline}) for $\gamma = -1/12$. The color code is as before.}
\label{fig3}
\end{figure}

Given the perfect agreement between the results obtained using the quasi pure approximation and those
from optimal witnesses, we now address a class of states which hitherto could not be characterized by 
the latter. 
These are states that extend ``beyond lines'' in 
the above sense, i.e., which cannot be generated by application of only one Weyl operator. We choose the 
following family:
\beq 
\rho =
(1-\alpha-\beta-\gamma)\frac{\openone}{9} +\gamma P_{00}+ \alpha
P_{10} +\beta P_{01} \; . \label{beyondline} 
\eeq 
We can once again easily scan the whole region of parameters for a fixed $\gamma$, and new areas of bound entanglement are found,  as illustrated in Fig.~\ref{fig3}.

%

 \begin{figure}[t!]
\includegraphics[width=\columnwidth]{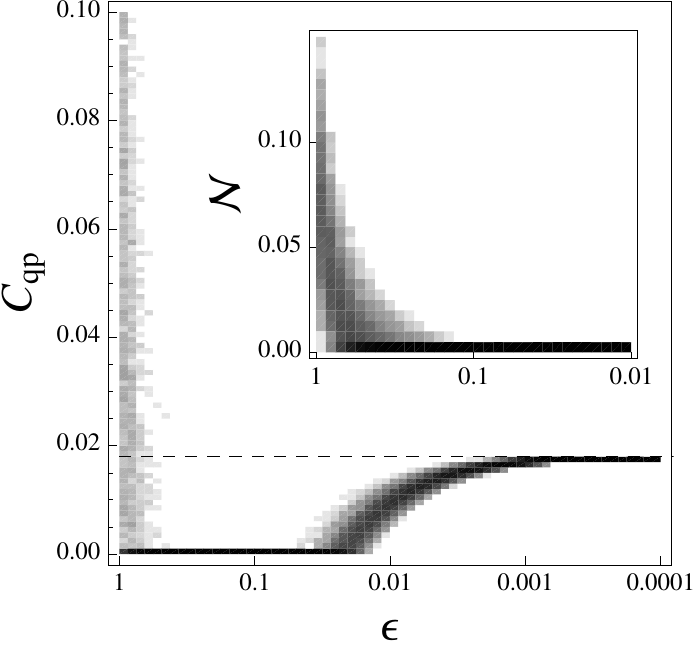}
\caption{Distribution of quasi pure concurrence $\mathcal{C}_{\rm qp}$, Eq.~(\ref{eq:lower_bound}), and of the negativity $\mathcal N$ (inset)
of $3\times 3$ random states, generated by admixture of  Hilbert-Schmidt-distributed random mixed states $\rho_{\rm HS}$ to the bound entangled 
reference ``line'' state $\rho_{\rm be}$, according to Eq.~(\ref{mixit}).
Saturation indicates the bin-populaton of the histogram for a fixed $\epsilon$.
}
\label{fig:volume}
\end{figure}

Our above results suggest that bound entangled states in general occupy a finite volume in the even higher dimensional state space of all $3 \times 3$ states,
as was actually proven in 
\cite{karol:883}, and underpinned for random bipartite states of dimension $2 \times 4$ in 
\cite{karol:3496}. In a similar vein as in~\cite{bandyopadhyay:032318}, we now show numerical evidence that our approach to detect bound entangled states is robust, i.e., it does not only work for Bell-diagonal states, but also for a finite bound entanglement \emph{volume} around them. As an example, we choose a  
``line-state" ($\rho_{\rm be}$) with
$\{\alpha,\beta,\gamma\} = \{-0.092, 0.04, 0.2148\}$ (and ${\cal
C}_{\rm qp} =0.018$) as defined in (\ref{online}) above, and mix it
with a  Hilbert-Schmidt-distributed random mixed state
($\rho_{\rm HS}$), as follows: 
\beq 
\tilde\rho=
(1-\epsilon)\rho_{\rm be}+\epsilon \,\rho_{\rm HS}, \mbox{\hspace{0.2cm}
with } 0\le \epsilon\le 1\, . 
\label{mixit}
\eeq 
In this way we are able to explicitly construct a ball of bound entangled states.

As before, the bound entangled fraction is identified by intersection of the area with 
positive partial transpose and that with non-vanishing quasi pure approximation $\mathcal{C}_{\rm qp}$. The result is
illustrated in Fig.~\ref{fig:volume} by the distribution of $\mathcal{C}_{\rm qp}$ and of the negativity $\mathcal N$~\cite{vidalNeg} (in the inset), as  function of the variable $\epsilon$. Similar plots are obtained for different initial bound entangled states, even for ``beyond line" states. Note that $\mathcal N$ vanishes if and only if the state has positive partial transpose, and can 
thus be used to demarcate the associated parameter range. 

For each value of $\epsilon$, $1000$ states $\rho_{\rm HS}$ were randomly chosen,
such that for $\epsilon=1$ we recover the entanglement characteristics of our sample -- where both, $\mathcal{C}_{\rm qp}$ and 
$\mathcal N$, exhibit a broad distribution. In the opposite limit, $\epsilon =0$, $\mathcal N$ vanishes identically, while 
$\mathcal{C}_{\rm qp}=0.018$ (identifed by the dashed line) -- indicating that $\rho_{\rm be}$ is bound entangled. The figure shows that 
already for $\epsilon \leq 0.01$ {\em all} sampled states have a positive partial transpose (vanishing $\mathcal N$) but non-vanishing 
$\mathcal{C}_{\rm qp}$, and are thus bound entangled. Since the states around $\rho_\text{be}$ are randomly chosen and are greater in number than the dimension of state space, the probability of all of them lying in a hyperplane is zero and thus they explore all directions in state space -- the convex hull of these points forms a body of finite volume in state space.  Therefore, our numerical result explicitly, and effortlessly, spots a finite volume of bound entanglement in the state space.  Nevertheless, we cannot rule out the existence of separable states in the constructed volume. But given the large number of our sample, and the fact that separable states form a convex set, the probability of such event is vanishingly small.

 {\it Acknowledgments.} The support by the DAAD-KRF GEnKO partnership (KRF-2009-614-C00001) is gladly acknowledge. F. de M. also acknowledges
the support by the Alexander von Humboldt Foundation. J. B. is supported by the IT R\&D program of MKE/IITA (2008-F-035-01).

\end{document}